# Energy-momentum Tensor for a Field and Particle in Interaction

Roderick I. Sutherland

Centre for Time, University of Sydney, NSW 2006 Australia

rod.sutherland@sydney.edu.au

## Abstract

This paper focuses on the basic system of a field and a particle in interaction and provides a single, unified derivation of the energy-momentum tensors for both the field and the particle. This derivation contrasts with the usual approach in text books, where these two tensors are derived separately via different considerations. The unified treatment thereby provides an immediate description of overall energy and momentum conservation for such a system.

## 1. Introduction

In describing a classical field interacting with a particle, it is common for text books to introduce a Lagrangian density expression from which one can derive the field equation but not the particle equation of motion. This expression then also allows derivation of the energy-momentum tensor for the field, but not for the particle (the latter being constructed via alternative procedures[1]). A few text books go further and present a unified Lagrangian density for the field and particle together, thereby allowing both the field and particle equations to be derived from the same expression. This more comprehensive approach prompts the thought that it should also be possible to carry out a single derivation for the overall energy-momentum tensor describing the combined field/particle system. Such a derivation, however, is not provided in any text known to the author. The subject of the present note is therefore to remedy this situation by detailing a suitable unified approach of this type.

## 2. Background information

A combined field/particle Lagrangian density usually takes the form:

$$\mathcal{L} = \mathcal{L}_{\text{field}} + \mathcal{L}_{\text{particle}} + \mathcal{L}_{\text{interaction}} \tag{1}$$

An example of such an expression is the electromagnetic Lagrangian density describing a particle of 4-velocity $u^\alpha$ interacting with a 4-potential $A^\alpha$. This is as follows[2]:

$$\mathcal{L} = -\tfrac{1}{4}F_{\alpha\beta}F^{\alpha\beta} - \sigma_0 m(u_\alpha u^\alpha)^{\frac{1}{2}} - \sigma_0 q\, u_\alpha A^\alpha \qquad (\alpha,\beta = 0,1,2,3) \tag{2}$$

---

[1] See, e.g., Sec. 1.6 in [1].

[2] See, e.g., Eq. (8.38) in [1], Eq. (13.125) in [2] and Eq. (6-100) in [3], although these are all presented in somewhat different form from Eq. (2) here.

Here $F^{\alpha\beta}$ is the electromagnetic field tensor, which can be expressed instead in terms of derivatives of $A^{\alpha}$, and m and q are the particle's mass and charge, respectively. The quantity $\sigma_0$ is the rest density distribution of the particle through space. This involves a delta function because, at any time, the particle's "matter density" is all concentrated at one point. The explicit form of $\sigma_0$ is:

$$\sigma_0 = \frac{1}{u^0}\delta^3[\mathbf{x} - \mathbf{x}_p(\tau)] \qquad (3)$$

where $\mathbf{x}_p$ is the particle's spatial position as a function of proper time $\tau$ along the particle's world line, and $\mathbf{x}$ is an arbitrary point in space. Eq. (2) has been written in manifestly Lorentz covariant form with metric tensor signature $+---$.

As noted in the introduction, Eq. (2) allows the derivation of not only the field equation but also the particle's equation of motion. These are found via the appropriate Euler-Lagrange equations which are generated from small variations in the field value and the particle's world line, respectively.

Another example of a combined field/particle Lagrangian density is the following one from which the well-known de Broglie-Bohm model for quantum mechanics can be derived [4,5]:

$$\mathcal{L} = \mathcal{L}_{\text{field}} - \sigma_0 b(u_\alpha u^\alpha)^{1/2} + \sigma_0 u_\alpha b^\alpha \qquad (4)$$

The form of this expression is seen to be similar to the electromagnetic case. Here $\mathcal{L}_{\text{field}}$ is the usual Lagrangian density corresponding to the field in question (e.g., Dirac or Klein-Gordon), $b^{\alpha}$ is a 4-vector potential which is a function of the quantum mechanical wavefunction and b is the magnitude of this 4-vector. The quantities $\sigma_0$ and $u^{\alpha}$ are again the particle's density distribution and 4-velocity, respectively. Once a statistical assumption is also included, it can be shown that expression (4) yields predictions consistent with quantum mechanics.

Note that, in deriving the equation of motion of a particle, it is actually a Lagrangian which is needed rather than a Lagrangian density. For both of the examples quoted above, the relationship between the particle Lagrangian L and the overall Lagrangian density $\mathcal{L}$ is:

$$\mathcal{L} = \mathcal{L}_{\text{field}} + \sigma_0 L \qquad (5)$$

This form is needed in order to be consistent with the corresponding expressions for action. In particular, the overall action S must be given by the following 4-dimensional integral:

$$S = \int \mathcal{L}\, d^4x \tag{6}$$

whereas the partial action relating to the particle needs to be related to the particle's proper time $\tau$ via the single integral:

$$S_p = \int L\, d\tau \tag{7}$$

The fact that inserting the particle term in Eq. (5) into the overall action (6) leads to the correct partial action (7) is shown in Appendix 1.

In the next section, a general expression for the energy-momentum tensor corresponding to a Lagrangian density of the relevant form (5) will be derived from the usual assumptions of symmetry under space and time displacements.

**3. Derivation of the energy-momentum tensor for a field/particle system**

It will be supposed that a particle of mass m and 4-velocity $u^\alpha$ is interacting with a complex scalar field $\phi$. A scalar field has been chosen here for simplicity because the further generalisation to a field with multiple components (such as the electromagnetic 4-potential discussed in Sec. 2) is straightforward. The combined field/particle Lagrangian density $\mathcal{L}$ is assumed to be an explicit function of all of the following: the field $\phi$ and its complex conjugate $\phi^*$, the first derivatives $\partial_\alpha \phi$ and $\partial_\alpha \phi^*$, the particle's rest density distribution in space $\sigma_0(x)$, the particle's rest mass m and 4-velocity $u^\alpha$ and perhaps the coordinates x:

$$\mathcal{L} = \mathcal{L}(\phi, \phi^*, \partial_\alpha \phi, \partial_\alpha \phi^*, \sigma_0, m, u^\alpha, x) \tag{8}$$

Now, Noether's theorem states that the system's energy and momentum will be conserved when $\mathcal{L}$ is symmetric under space and time displacements, i.e., when it is not an explicit function of the coordinates x. This symmetry means it is sufficient to write:

$$\mathcal{L} = \mathcal{L}(\phi, \phi^*, \partial_\alpha \phi, \partial_\alpha \phi^*, \sigma_0, m, u^\alpha) \tag{9}$$

where the x has been deleted. The dependence of $\mathcal{L}$ on the particle's rest mass and 4-velocity will also turn out to be irrelevant here, because m is a constant and $u^\alpha$ is not a function of the coordinates (it is only a function of the proper time $\tau$ along the particle's world line). Therefore, instead of Eq. (9), the relevant dependence for present purposes can be written in the reduced form:

$$\mathcal{L} = \mathcal{L}(\phi, \phi^*, \partial_\alpha \phi, \partial_\alpha \phi^*, \sigma_0) \tag{10}$$

The partial derivative of $\mathcal{L}$ with respect to the particular coordinate $x^\mu$ will now be taken, holding the other three coordinates $x^\beta$ constant, where $\beta \neq \mu$. The quantities held constant in each partial differentiation will initially be shown explicitly outside square brackets for greater clarity. From Eq. (10), the full expression for this derivative is then:

$$\left[\frac{\partial \mathcal{L}}{\partial x_\mu}\right]_{x_\beta} = \left[\frac{\partial \mathcal{L}}{\partial \phi}\right]_{\phi^*,\partial_\alpha\phi,\partial_\alpha\phi^*,\sigma_0}\left[\frac{\partial \phi}{\partial x_\mu}\right]_{x_\beta} + \left[\frac{\partial \mathcal{L}}{\partial \phi^*}\right]_{\phi,\partial_\alpha\phi,\partial_\alpha\phi^*,\sigma_0}\left[\frac{\partial \phi^*}{\partial x_\mu}\right]_{x_\beta}$$
$$+ \left[\frac{\partial \mathcal{L}}{\partial(\partial_\nu\phi)}\right]_{\phi,\phi^*,\partial_\alpha\phi^*,\sigma_0}\left[\frac{\partial(\partial_\nu\phi)}{\partial x_\mu}\right]_{x_\beta} + \left[\frac{\partial \mathcal{L}}{\partial(\partial_\nu\phi^*)}\right]_{\phi,\phi^*,\partial_\alpha\phi,\sigma_0}\left[\frac{\partial(\partial_\nu\phi^*)}{\partial x_\mu}\right]_{x_\beta}$$
$$+ \left[\frac{\partial \mathcal{L}}{\partial \sigma_0}\right]_{\phi,\phi^*,\partial_\alpha\phi,\partial_\alpha\phi^*}\left[\frac{\partial \sigma_0}{\partial x_\mu}\right]_{x_\beta} \tag{11}$$

$(\mu, \nu, \alpha, \beta = 0,1,2,3)$

The third and fourth terms on the right in Eq. (11) each contain a summation over $\nu$. Note that in the less simple case of Eq. (8), where $\mathcal{L}$ is also a function of the coordinates, the right hand side of Eq. (11) would contain an extra term of the form:

$$\left[\frac{\partial \mathcal{L}}{\partial x_\mu}\right]_{\phi,\phi^*,\partial_\alpha\phi,\partial_\alpha\phi^*,\sigma_0,x_\beta} \tag{12}$$

since this term would not be zero. Also note how this term differs from the one on the left hand side of Eq. (11).

From here on, the quantities held constant will no longer be shown. Eq. (11) is then written more neatly as:

$$\frac{\partial \mathcal{L}}{\partial x_\mu} = \frac{\partial \mathcal{L}}{\partial \phi}\partial^\mu\phi + \frac{\partial \mathcal{L}}{\partial \phi^*}\partial^\mu\phi^* + \frac{\partial \mathcal{L}}{\partial(\partial_\nu\phi)}\partial^\mu(\partial_\nu\phi) + \frac{\partial \mathcal{L}}{\partial(\partial_\nu\phi^*)}\partial^\mu(\partial_\nu\phi^*) + \frac{\partial \mathcal{L}}{\partial \sigma_0}\partial^\mu\sigma_0 \tag{13}$$

Now the first and second terms on the right of Eq. (13) can be modified using the field equations for $\phi$ and $\phi^*$. From standard Lagrangian theory, these equations are [2]:

$$\partial_\nu \frac{\partial \mathcal{L}}{\partial(\partial_\nu\phi)} = \frac{\partial \mathcal{L}}{\partial \phi} \tag{14}$$

$$\partial_\nu \frac{\partial \mathcal{L}}{\partial(\partial_\nu\phi^*)} = \frac{\partial \mathcal{L}}{\partial \phi^*} \tag{15}$$

and (13) can therefore be written in the form:

$$\frac{\partial \mathcal{L}}{\partial x_\mu} = \left\{ \partial_\nu \frac{\partial \mathcal{L}}{\partial(\partial_\nu \phi)} \right\} \partial^\mu \phi + \left\{ \partial_\nu \frac{\partial \mathcal{L}}{\partial(\partial_\nu \phi^*)} \right\} \partial^\mu \phi^* + \frac{\partial \mathcal{L}}{\partial(\partial_\nu \phi)} \partial^\mu \partial_\nu \phi + \frac{\partial \mathcal{L}}{\partial(\partial_\nu \phi^*)} \partial^\mu \partial_\nu \phi^* + \frac{\partial \mathcal{L}}{\partial \sigma_0} \partial^\mu \sigma_0 \tag{16}$$

This equation can be manipulated further as follows:

$$\begin{aligned} \frac{\partial \mathcal{L}}{\partial x_\mu} &= \partial_\nu \left\{ \frac{\partial \mathcal{L}}{\partial(\partial_\nu \phi)} \partial^\mu \phi + \frac{\partial \mathcal{L}}{\partial(\partial_\nu \phi^*)} \partial^\mu \phi^* \right\} + \frac{\partial \mathcal{L}}{\partial \sigma_0} \partial^\mu \sigma_0 \\ &= \partial_\nu \left\{ \frac{\partial \mathcal{L}}{\partial(\partial_\nu \phi)} \partial^\mu \phi + \frac{\partial \mathcal{L}}{\partial(\partial_\nu \phi^*)} \partial^\mu \phi^* \right\} + \partial^\mu \left( \frac{\partial \mathcal{L}}{\partial \sigma_0} \sigma_0 \right) - \sigma_0 \partial^\mu \left( \frac{\partial \mathcal{L}}{\partial \sigma_0} \right) \\ &= \partial_\nu \left\{ (\partial^\mu \phi) \frac{\partial \mathcal{L}}{\partial(\partial_\nu \phi)} + (\partial^\mu \phi^*) \frac{\partial \mathcal{L}}{\partial(\partial_\nu \phi^*)} + g^{\mu\nu} \frac{\partial \mathcal{L}}{\partial \sigma_0} \sigma_0 \right\} - \sigma_0 \partial^\mu \left( \frac{\partial \mathcal{L}}{\partial \sigma_0} \right) \end{aligned} \tag{17}$$

Now, as pointed out earlier in Eq. (5), a combined Lagrangian density for a field and particle in interaction will have the general form:

$$\mathcal{L} = \mathcal{L}_{\text{field}} + \sigma_0 L \tag{18}$$

where L is the Lagrangian for the particle. Since $\mathcal{L}_{\text{field}}$ and L are not functions of $\sigma_0$, expression (18) allows (17) to be written as:

$$\frac{\partial \mathcal{L}}{\partial x_\mu} = \partial_\nu \left\{ (\partial^\mu \phi) \frac{\partial \mathcal{L}}{\partial(\partial_\nu \phi)} + (\partial^\mu \phi^*) \frac{\partial \mathcal{L}}{\partial(\partial_\nu \phi^*)} + g^{\mu\nu} L \sigma_0 \right\} - \sigma_0 \partial^\mu L \tag{19}$$

As shown in Appendix 2, the last term in this equation can be modified using the equation of motion for the particle so that Eq. (19) then becomes:

$$\frac{\partial \mathcal{L}}{\partial x_\mu} = \partial_\nu \left\{ (\partial^\mu \phi) \frac{\partial \mathcal{L}}{\partial(\partial_\nu \phi)} + (\partial^\mu \phi^*) \frac{\partial \mathcal{L}}{\partial(\partial_\nu \phi^*)} + g^{\mu\nu} L \sigma_0 + \sigma_0 p^\mu u^\nu \right\} \tag{20}$$

where $p^\mu$ is the particle's generalised momentum. The left and right hand sides of this equation can now be combined to give:

$$\partial_\nu \left\{ (\partial^\mu \phi) \frac{\partial \mathcal{L}}{\partial(\partial_\nu \phi)} + (\partial^\mu \phi^*) \frac{\partial \mathcal{L}}{\partial(\partial_\nu \phi^*)} + g^{\mu\nu} L \sigma_0 + \sigma_0 p^\mu u^\nu - g^{\mu\nu} \mathcal{L} \right\} = 0 \tag{21}$$

Using (18) again then finally yields:

$$\partial_\nu \left\{ (\partial^\mu \phi) \frac{\partial \mathcal{L}}{\partial(\partial_\nu \phi)} + (\partial^\mu \phi^*) \frac{\partial \mathcal{L}}{\partial(\partial_\nu \phi^*)} - g^{\mu\nu} \mathcal{L}_{\text{field}} + \sigma_0 p^\mu u^\nu \right\} = 0 \tag{22}$$

Eq. (22) shows that the expression in curly brackets has zero 4-divergence, which allows it to be identified as the overall energy-momentum tensor $T^{\mu\nu}$ for the combined field and particle system:

$$T^{\mu\nu} = (\partial^{\mu}\phi)\frac{\partial \mathcal{L}}{\partial(\partial_{\nu}\phi)} + (\partial^{\mu}\phi^{*})\frac{\partial \mathcal{L}}{\partial(\partial_{\nu}\phi^{*})} - g^{\mu\nu}\mathcal{L}_{\text{field}} + \sigma_0 p^{\mu}u^{\nu} \tag{23}$$

A general expression for describing conservation of energy and momentum in the system has therefore been obtained[3].

It is also instructive to separate this expression into three parts:

$$T^{\mu\nu} = T^{\mu\nu}_{\text{field}} + T^{\mu\nu}_{\text{particle}} + T^{\mu\nu}_{\text{interaction}} \tag{24}$$

where the individual terms are defined to be:

$$T^{\mu\nu}_{\text{field}} = (\partial^{\mu}\phi)\frac{\partial \mathcal{L}_{\text{field}}}{\partial(\partial_{\nu}\phi)} + (\partial^{\mu}\phi^{*})\frac{\partial \mathcal{L}_{\text{field}}}{\partial(\partial_{\nu}\phi^{*})} - g^{\mu\nu}\mathcal{L}_{\text{field}} \tag{25}$$

$$T^{\mu\nu}_{\text{particle}} = \sigma_0 p^{\mu}u^{\nu} \tag{26}$$

$$T^{\mu\nu}_{\text{interaction}} = \sigma_0(\partial^{\mu}\phi)\frac{\partial L}{\partial(\partial_{\nu}\phi)} + \sigma_0(\partial^{\mu}\phi^{*})\frac{\partial L}{\partial(\partial_{\nu}\phi^{*})} \tag{27}$$

Here Eq. (25) is seen to be the commonly quoted expression for a free field existing in the absence of particles [2]. In the more general case where Eqs. (26) and (27) are involved as well, the 4-divergences of the three terms need not be separately zero. Conservation for the system as a whole, however, requires that the total 4-divergence must be zero, a mathematical condition which holds here.

Eqs. (25) to (27) taken together thus provide the desired, overall energy-momentum tensor associated with a field and a particle in interaction[4], with the separate contributions for the field and the particle having been found via a single, unified derivation.

## Appendix 1

The aim here is to confirm that inserting the particle term in Eq. (5) into the overall action expression in Eq. (6) leads to the correct partial action as given in Eq. (7). Substituting the last term of (5) into (6) yields:

---

[3] The overall tensor $T^{\mu\nu}$ defined here is actually the "canonical" energy-momentum tensor, which is not necessarily symmetric and hence does not necessarily conserve angular momentum. Techniques exist to symmetrise this tensor [6].

[4] In the particular case of electromagnetism, the term $T^{\mu\nu}_{\text{interaction}}$ is found to be zero.

$$S_p = \int \sigma_0 L\, d^4x \tag{28}$$

Using Eq. (3) this becomes:

$$\begin{aligned} S_p &= \int \frac{1}{u^0} \delta^3[\mathbf{x} - \mathbf{x}_p(\tau)] L\, d^4x \\ &= \int \frac{1}{u^0} L\, dx^0 \end{aligned} \tag{29}$$

Employing the definition of 4-velocity:

$$u^\alpha = \frac{dx^\alpha}{d\tau} \tag{30}$$

then allows Eq. (29) to be re-expressed as:

$$\begin{aligned} S_p &= \int \frac{d\tau}{dx^0} L\, dx^0 \\ &= \int L\, d\tau \end{aligned} \tag{31}$$

which is Eq. (7) as required.

## Appendix 2

The steps which lead from Eq. (19) to Eq. (20) will be presented here. Standard Lagrangian theory [2] provides the following equation of motion for the particle[5]:

$$\frac{dp^\mu}{d\tau} = -\partial^\mu L \tag{32}$$

where the generalised momentum $p^\mu$ of the particle is defined by:

$$p^\mu \equiv -\frac{\partial L}{\partial u_\mu} \tag{33}$$

The last term in Eq. (19) can then be written as:

---

[5] The negative signs in both Eq. (32) and Eq. (33) have been chosen to be consistent with the more familiar, non-relativistic equations $dp^i / dt = \partial L / \partial x^i$ and $p^i = \partial L / \partial v^i$. (where $i = 1,2,3$ and $x^i = -x_i$).

$$
\begin{aligned}
-\sigma_0 \partial^\mu L &= \sigma_0 \frac{dp^\mu}{d\tau} \\
&= \sigma_0 \frac{dx^\nu}{d\tau} \frac{\partial p^\mu}{\partial x^\nu} \\
&= \sigma_0 u^\nu \partial_\nu p^\mu \\
&= \partial_\nu (\sigma_0 u^\nu p^\mu) - p^\mu \partial_\nu (\sigma_0 u^\nu)
\end{aligned}
\tag{34}
$$

Now, referring back to Sec. 2, the quantity $\sigma_0$ represents the spatial distribution (here a delta function) of the "matter" making up the particle. Taking the particle's matter density to be conserved, the following continuity equation will be satisfied:

$$
\partial_\nu (\sigma_0 u^\nu) = 0 \tag{35}
$$

This means that Eq. (34) reduces to:

$$
-\sigma_0 \partial^\mu L = \partial_\nu (\sigma_0 u^\nu p^\mu) \tag{36}
$$

which allows Eq. (19) to be expressed in the form of Eq. (20), as required.